\documentclass[aps,a4paper,10pt,superscriptaddress,twocolumn,prl]{revtex4-2}
\usepackage[utf8]{inputenc}
\usepackage{amsmath}
\usepackage{amssymb}
\usepackage{braket}	
\usepackage{graphicx}
\usepackage{xcolor}
\usepackage{mathrsfs}
\usepackage[colorlinks=true, citecolor=.]{hyperref}
\usepackage{soul}
\usepackage{physics}

\begin{document}

\title{Beating three-parameter precision trade-offs with entangling collective measurements}

\date{\today}

\author{Simon K. Yung}
\thanks{These authors contributed equally to this work}
\affiliation{Department of Quantum Science and Technology, Research School of Physics, The Australian National University, Canberra, ACT 2601, Australia.}

\author{Wen-Zhe Yan}
\thanks{These authors contributed equally to this work}
\affiliation{Laboratory of Quantum Information, University of Science and Technology of China, Hefei, 230026, China}
\affiliation{Anhui Province Key Laboratory of Quantum Network, University of Science and Technology of China, Hefei, 230026, China}

\author{Lan-Tian Feng}
\email{fenglt@ustc.edu.cn}
\affiliation{Laboratory of Quantum Information, University of Science and Technology of China, Hefei, 230026, China}
\affiliation{Anhui Province Key Laboratory of Quantum Network, University of Science and Technology of China, Hefei, 230026, China}
\affiliation{CAS Center For Excellence in Quantum Information and Quantum Physics, University of Science and Technology of China, Hefei, 230026,  China}
\affiliation{Hefei National Laboratory, Hefei, 230088, China}

\author{Aritra Das}
\affiliation{Department of Quantum Science and Technology, Research School of Physics, The Australian National University, Canberra, ACT 2601, Australia.}

\author{Jiayi Qin}
\affiliation{Department of Quantum Science and Technology, Research School of Physics, The Australian National University, Canberra, ACT 2601, Australia.}

\author{Guang-Can Guo}
\affiliation{Laboratory of Quantum Information, University of Science and Technology of China, Hefei, 230026, China}
\affiliation{Anhui Province Key Laboratory of Quantum Network, University of Science and Technology of China, Hefei, 230026, China}
\affiliation{CAS Center For Excellence in Quantum Information and Quantum Physics, University of Science and Technology of China, Hefei, 230026,  China}
\affiliation{Hefei National Laboratory, Hefei, 230088, China}

\author{Ping Koy Lam}
\affiliation{A*STAR Quantum Innovation Centre (Q.INC), Agency for Science, Technology and Research (A*STAR), 2 Fusionopolis Way, Innovis, 138634, Singapore.}
\affiliation{Centre for Quantum Technologies, National University of Singapore, 3 Science Drive 2, Singapore 117543, Singapore.}

\author{Jie Zhao}
\email{jie.zhao@anu.edu.au}
\affiliation{Department of Quantum Science and Technology, Research School of Physics, The Australian National University, Canberra, ACT 2601, Australia.}

\author{Zhibo Hou}
\email{houzhibo@ustc.edu.cn}
\affiliation{Laboratory of Quantum Information, University of Science and Technology of China, Hefei, 230026, China}
\affiliation{Anhui Province Key Laboratory of Quantum Network, University of Science and Technology of China, Hefei, 230026, China}
\affiliation{CAS Center For Excellence in Quantum Information and Quantum Physics, University of Science and Technology of China, Hefei, 230026,  China}
\affiliation{Hefei National Laboratory, Hefei, 230088, China}

\author{Lorc\'an O. Conlon}
\affiliation{A*STAR Quantum Innovation Centre (Q.INC), Agency for Science, Technology and Research (A*STAR), 2 Fusionopolis Way, Innovis, 138634, Singapore.}
\affiliation{Centre for Quantum Technologies, National University of Singapore, 3 Science Drive 2, Singapore 117543, Singapore.}

\author{Syed M. Assad}
\email{cqtsma@gmail.com}
\affiliation{A*STAR Quantum Innovation Centre (Q.INC), Agency for Science, Technology and Research (A*STAR), 2 Fusionopolis Way, Innovis, 138634, Singapore.}

\author{Xi-Feng Ren}
\affiliation{Laboratory of Quantum Information, University of Science and Technology of China, Hefei, 230026, China}
\affiliation{Anhui Province Key Laboratory of Quantum Network, University of Science and Technology of China, Hefei, 230026, China}
\affiliation{CAS Center For Excellence in Quantum Information and Quantum Physics, University of Science and Technology of China, Hefei, 230026,  China}
\affiliation{Hefei National Laboratory, Hefei, 230088, China}

\author{Guo-Yong Xiang}
\affiliation{Laboratory of Quantum Information, University of Science and Technology of China, Hefei, 230026, China}
\affiliation{Anhui Province Key Laboratory of Quantum Network, University of Science and Technology of China, Hefei, 230026, China}
\affiliation{CAS Center For Excellence in Quantum Information and Quantum Physics, University of Science and Technology of China, Hefei, 230026,  China}
\affiliation{Hefei National Laboratory, Hefei, 230088, China}

\begin{abstract}
    Quantum-mechanical incompatibility, which precludes the simultaneous precise measurement of non-commuting observables, imposes fundamental limits on the rate at which classical information can be extracted. While the potential to surpass these limits using entangling collective measurements has been explored for two parameters, the regime of three or more parameters remains largely unexplored despite its fundamental and technological importance. Here, we investigate the three-parameter trade-off relations for estimating the Bloch vector components of a qubit, comparing conventional individual measurements with entangling collective measurements. We theoretically derive and experimentally implement optimal collective measurements on two identically prepared qubits using a programmable photonic circuit. Our experimental results demonstrate a clear violation of the entanglement-free trade-off relation—by an average of 16 standard deviations—achieving a tomography precision beyond the reach of any individual measurement scheme. This work directly confirms that optimal collective measurements can surpass the fundamental quantum limits of individual schemes in a three-parameter setting—thereby deepening our understanding of quantum uncertainty relations beyond the two-parameter regime and providing a clear strategy to overcome the precision trade-offs imposed by quantum incompatibility.

\end{abstract}

\maketitle

\textit{Introduction}---Quantum mechanics sets a fundamental barrier to classical information extraction: non-commuting observables cannot be measured simultaneously with arbitrary precision~\cite{heisenberg_uber_1927}. This incompatibility manifests as an inescapable trade-off ~\cite{liu_quantum_2019,albarelli_perspective_2020,demkowicz-dobrzanski_multi-parameter_2020,ragy_compatibility_2016} in the estimation of multiple parameters encoded in a quantum state—a core challenge across quantum sensing~\cite{giovannetti_quantum_2006,taylor_high-sensitivity_2008,caves_quantum-mechanical_1981}, quantum control and feedback~\cite{robertson_uncertainty_1929,schrodinger_zum_1930} and quantum communication~\cite{pirandola_advances_2020}. 

While this trade-off is well-characterized for two parameters in terms of variance-based~\cite{robertson_uncertainty_1929,schrodinger_zum_1930}, entropic~\cite{bialynicki-birula_uncertainty_1975}, majorization~\cite{partovi_majorization_2011}, and error-disturbance-based formulations~\cite{ozawa_universally_2003,busch_proof_2013,branciard_error-tradeoff_2013}, a profound gap exists for three or more.  The pairwise nature of commutators means that simple combinations of two-parameter limits fail to capture the complex, multi-dimensional constraints at play, and does not generally yield attainable bounds~\cite{robertson_indeterminacy_1934}. 

Recent experiments have demonstrated that collective measurements on multiple copies of a state can enhance two-parameter estimation precision~\cite{vidrighin_joint_2014,hou_deterministic_2018,roccia_entangling_2018,conlon_approaching_2023,conlon_discriminating_2023,zhou_experimental_2025}, but have not investigated three-parameter trade-offs. Whether the ultimate trade-off surface for three parameters can be saturated by collective strategies remains an open question, even in the simplest quantum systems. 

Here, we address this question by studying trade-offs in qubit tomography under collective measurements on two copies. In addition to the fundamental nature of this task, qubit tomography is an essential primitive in many emerging quantum technologies~\cite{gaikwad_gradient-descent_2025,hansenne_optimal_2025,acharya_pauli_2025,huang_predicting_2020,wootters_optimal_1989,rehacek_minimal_2004,scott_tight_2006,petz_efficient_2012,hou_achieving_2016,li_optimal_2023}. We determine the fundamental trade-off using Cram\'er--Rao-type bounds. We then construct an optimal two-copy measurement and implement it on a programmable photonic chip. Our results demonstrate that the fundamental precision trade-off surface can be saturated, strictly outperforming all single-copy strategies. 
Our findings establish a new benchmark for multiparameter quantum estimation and offer practical insights for quantum tomography, calibration, and quantum-enhanced sensing in systems constrained by incompatible observables.

\begin{figure*}
	\centering
	\includegraphics[width=\linewidth]{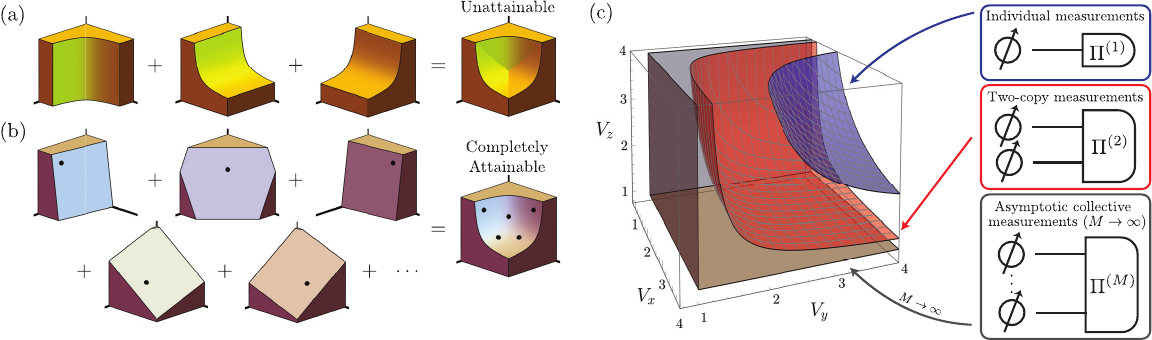}
	\caption{(a) Two-parameter trade-off relations for each pair of parameters can be combined to obtain a three-parameter trade-off region bounded by a trade-off surface. The overall unfilled region is not necessarily attainable, but the filled region is forbidden. (b) A Cram\'er--Rao lower bound forbids a half-space of the error-space, bounded by a plane determined by the weights. The combination of all such lower bounds can be attainable when each weighted bound is attainable; the resulting surface represents the fundamental limit. The black dot on each constituent surface is attainable.  (c) Trade-off surface for qubit tomography with individual measurements (blue), two-copy measurements (red), and asymptotic collective measurements (grey).}
	\label{fig:conceptual}
\end{figure*}

\textit{Trade-off relations for three-parameter estimation}---We study the practical task of qubit state tomography through the lens of quantum multiparameter estimation. Writing the state of a single qubit as $\rho_\theta = (\openone_2 + \theta\cdot\sigma)/2$, where $\sigma=(\sigma_x,\sigma_y,\sigma_z)$ are the Pauli operators, the state of the qubit is determined by the values of the real parameters $\theta=(\theta_x,\theta_y,\theta_z)$. These parameters can be estimated by measuring identically prepared copies of the quantum state with a measurement described by a positive operator-valued measure (POVM). As is common in the literature, we work in a local framework where we assume that the parameters are approximately known, $\theta\approx \theta_0$, and we consider sensing small changes in the parameters around $\theta_0$. This is a reasonable assumption when the number of available copies is large~\cite{barndorff-nielsen_fisher_2000,hayashi_statistical_2005,paris_quantum_2009,holevo_probabilistic_2011}.

The goal here is to minimize the error of the estimates, $(\hat\theta_x,\hat\theta_y,\hat\theta_z)$, as measured by their mean squared errors, $(V_x,V_y,V_z)$, which are the diagonal elements of the mean squared error matrix with elements $V_{ij}=N\mathbb{E}[(\theta_i-\hat\theta_i)(\theta_j-\hat\theta_j)]$, where $N$ is the number of copies of the state used and is included for normalization---in this way, we study the efficiency of the estimation process. The mean squared errors coincide with the estimation variances when the estimates are unbiased. Due to the incompatibility of the optimal measurements for the different parameters, it is generally impossible to attain the global minima of each variance simultaneously. This raises the questions: what is the trade-off between the minimum attainable variances, how does this trade-off change when collective measurements are allowed, and how can these limits be saturated?

Existing trade-off relations are exclusively applicable for two-parameter estimation~\cite{lu_incorporating_2021,yung_comparison_2024}. For qubit tomography with individual measurements, this gives (forming a two-parameter model by treating the third parameter as known) $(V_i-a)(V_j-b)\geq c$ for $i\neq j \in\{x,y,z\}$ and $a,b,c$ determined by $\theta_0$~\cite{yung_comparison_2024}. This reflects that only a finite amount of information can be extracted from the system, and reducing the error of one parameter incurs a penalty on the other. We could obtain a three-parameter trade-off relation by combining the two-parameter relations for each pair of parameters, as depicted in Fig.~\ref{fig:conceptual}(a). 
The resulting trade-off relation is valid, but crucially, is only asymptotically saturable because each two-parameter relation is attainable only in the limit where $V_k\rightarrow \infty$ ($k\neq i,j$). 

In this work, we alleviate this restriction, presenting trade-off relations that account for incompatibilities inherent in simultaneously estimating three parameters. We use Cram\'er--Rao-type lower bounds that restrict variances as $\Tr[WV] = \sum_{i\in\{x,y,z\}} w_i V_i \geq \mathcal{C}(W)$, where $w_i$ are called weights and are the elements of the diagonal positive-semidefinite weight matrix $W$, and $\mathcal{C}(W)$ is a lower bound. We use the Nagaoka--Hayashi Cram\'er--Rao bound (NHCRB)~\cite{nagaoka_new_2005,nagaoka_generalization_2005,conlon_efficient_2021} which, as we will see, is attainable for this system but applies only to individual measurements. We extend its validity by considering two-copy measurements as individual measurements on $\rho_\theta^{\otimes 2} = \rho_\theta\otimes \rho_\theta$. 

For simplicity, the results we present here are for $\theta_0=(0,0,0)$, corresponding to the maximally-mixed qubit state. Equivalent results can be obtained for any $\theta_0$, as we demonstrate in the Supplemental Material. In this case, the lower bounds for the mean squared errors with single-copy and two-copy measurements are
\begin{equation}
    \textstyle\sum_{i} w_i V_i\geq \mathcal{C}^{(1)}(W) = \left(\textstyle\sum_{i}\sqrt{w_i}\right)^2 \label{eq:NH1}
\end{equation}
and
\begin{equation}
    \textstyle\sum_{i} w_i V_i\geq \mathcal{C}^{(2)}(W) = \textstyle\sum_{i} w_i + \textstyle\sum_{i\neq j}\sqrt{w_iw_j},\label{eq:NH2} 
\end{equation}
respectively, where $i,j\in\{x,y,z\}$ (see Supplemental Material for details on the calculation of the bounds). The difference is greatest at equal weights, indicating that collective measurements are most beneficial when estimating the parameters equally. On the other hand, collective measurements offer no advantage for single-parameter estimation (the limit in which two weights are zero).

\begin{figure*}
	\centering
	\includegraphics[width=\linewidth]{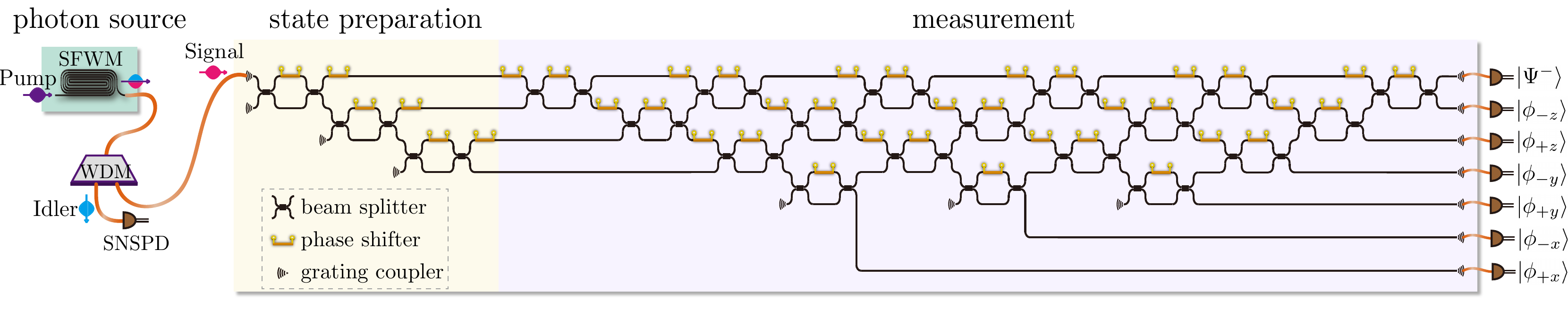}
	\caption{Experimental setup based on silicon integrated optics. A 1561.42 nm herald single photon generated by an integrated spontaneous four-wave mixing (SFWM) source is sent to a programmable photonic chip that integrates both the state preparation and measurement modules.
    The photon can be prepared in an arbitrary path-encoded four-dimensional pure state using three Mach-Zehnder interferometers (MZIs), each comprising two beams splitters and two reconfigurable phase shifters. Then, the measurement module, built from a cascaded MZI network, performs the optimal seven-outcome two-qubit measurement on the input state.  
    WDM, wavelength division multiplexing; SNSPD, superconducting nanowire single-photon detector.
    }
	\label{fig:exp}
\end{figure*}

For each set of weights, the lower bound Eq.~\eqref{eq:NH1} or Eq.~\eqref{eq:NH2} forbids a half-space of the error-space. An overall region can be determined by considering all possible weights~\cite{kull_uncertainty_2020,assad_accessible_2020}, see Fig.~\ref{fig:conceptual}(b), and will be attainable if the underlying Cram\'er--Rao bound is attainable. Each set of weights contributes a point to the boundary of the overall region (black points in Fig.~\ref{fig:conceptual}(b)); correspondingly, the weighted trace is a useful scalar quantity to assess proximity to the trade-off surface. We determine the trade-off relations analytically (see Supplemental Material), and find for single-copy measurements
\begin{equation}
	V_xV_yV_z -V_xV_y-V_xV_z-V_yV_z \geq 0, \quad V_i\geq 1. \label{eq:tradeoff1}
\end{equation}
Similarly, the trade-off relation for two-copy measurements is  
\begin{equation}
V_xV_yV_z -V_xV_y-V_xV_z-V_yV_z + \tfrac{3}{4}(V_x+V_y+V_z)\geq \tfrac{1}{2}	\label{eq:tradeoff2}
\end{equation}
and $V_i\geq 1$. Equality in these relations corresponds to trade-off surfaces that represent the fundamental limit for the mean squared errors of single- and two-copy measurements. The surfaces are presented in Fig.~\ref{fig:conceptual}(c), where the collective advantage is evident. In the limit of an asymptotically large collective measurement, the trade-off can be entirely eliminated (see Supplementary Material).  

\textit{Surpassing the single-copy trade-off relation}---When restricted to measurements performed only on single qubits, an optimal strategy is to measure $\sigma_i$ with probability $p_i$~\cite{hou_achieving_2016}. For every $\{p_i\}$, the measurement attains the trade-off relation in Eq.~\eqref{eq:tradeoff1}, with the position on the surface dictated by the probabilities (see Supplementary Material).

To beat the single-copy limit, entanglement at the measurement stage is necessary. We present an optimal POVM for the maximally-mixed state consisting of projectors onto entangled states in each Pauli basis, plus a singlet state: $\Pi^{(2)} = \{\ketbra{\phi_{+i}},\ketbra{\phi_{-i}},\ketbra{\Psi^-}; \ i=x,y,z\}$. The states are of the form $\ket{\phi_{\pm i}}=\alpha_{\pm i}\ket{0_i}\ket{0_i} + \alpha_{\mp i} \ket{1_i}\ket{1_i}$, where $\ket{0_i}$ and $\ket{1_i}$ are the eigenstates of $\sigma_i$, and {$\ket{\Psi^-}=\left( \ket{0_z}\ket{1_z}-\ket{1_z}\ket{0_z} \right)/\sqrt{2}$}. The trade-off surface can be attained by selecting the coefficients $\alpha_{\pm i}$, which are defined by the weights for which the measurement is optimal, and are detailed in the End Matter. As a result, the collective measurement $\Pi^{(2)}$ surpasses the single-copy limit and saturates any point on the two-copy trade-off surface.   

Similarly to the single-copy measurement, the first 3 pairs of elements of $\Pi^{(2)}$ exclusively give information about one of the three parameters, although utilising entanglement whenever the weights are not equal. It is also noteworthy that the two-copy optimal measurement is a genuine simultaneous measurement of the three parameters. This is in contrast to the single-copy measurement which can also be expressed as three separate measurements. That is, in the single-copy case one could simply measure $\sigma_{x(y,z)}$ on a fraction of the probes, whereas in the two-copy case, all probes should be measured pairwise with $\Pi^{(2)}$.

\begin{figure*}
	\centering
	\includegraphics[width=0.8\linewidth]{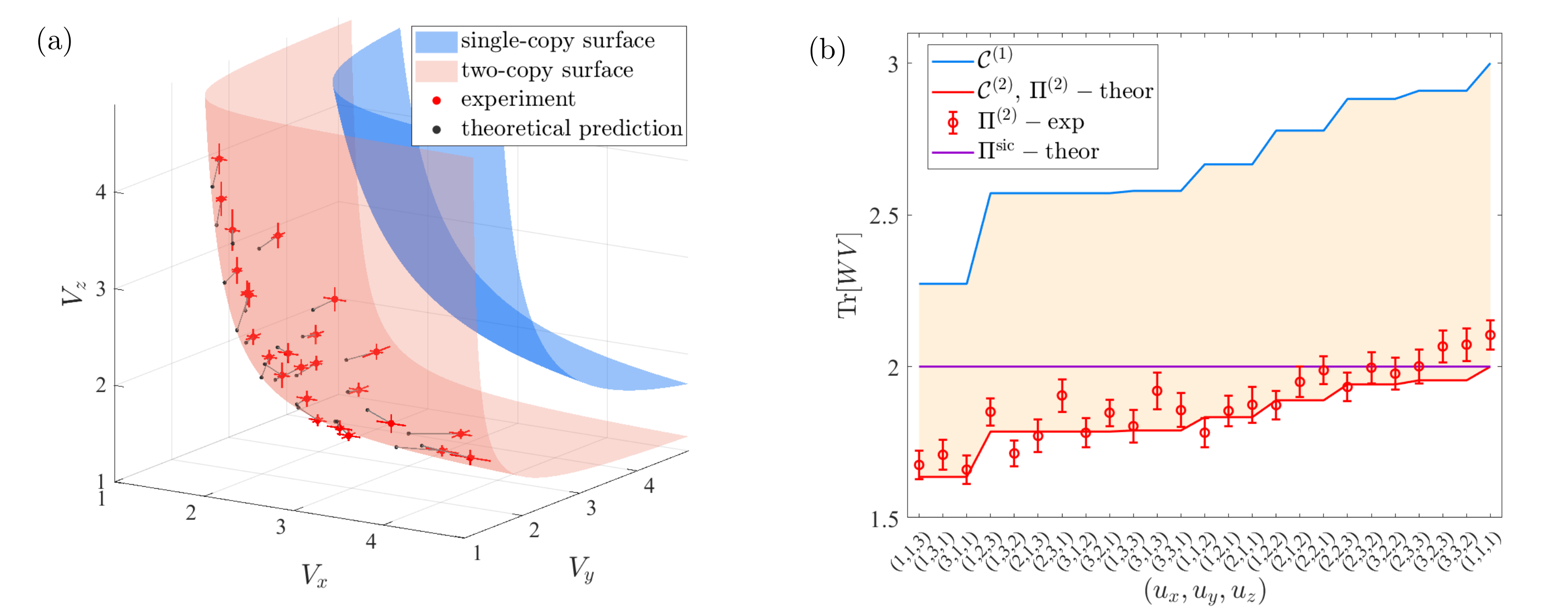}
	\caption{Experimental results for the optimal two-copy qubit state tomography with true parameters $\theta_x=\theta_y=\theta_z=0$.
    (a) Three-parameter trade-off surface. The blue and red surfaces show the estimation limits when using individual and two-copy collective measurements, respectively. Each experimental data point is obtained using the optimal measurement for a specific choice of the weight matrix. 
    (b) Weighted traces of the mean squared error matrix, representing the total estimation error. 
    The horizontal axis corresponds to the weight matrix via $W= \operatorname{Diag}(u_x^2,u_y^2,u_z^2)/(u_x^2+u_y^2+u_z^2)$. The two-copy experimental estimation errors (red circles) agree well with the theoretical results (red curve, from Eq.~\eqref{eq:NH2}). The performance of the two-copy SIC-POVM $\Pi^\text{sic}$ is also presented for comparison. The gap between the single-copy limit (blue curve) and two-copy results showcases the advantage of collective measurements. The error bars denote the standard deviations of 50 numerical simulations from Poisson statistics. 
    }
	\label{fig:results_0}
\end{figure*}

\textit{Experimental results}---We demonstrate the optimal two-copy qubit tomography protocol using a programmable photonic chip illustrated in Fig.~\ref{fig:exp}.
On this platform, quantum states are encoded in the path degree of freedom of a single photon, with four path modes spanning the four-dimensional Hilbert space of two qubits.
The manipulation of the path state is realized by combinations of Mach-Zehnder interferometers (MZIs). Each MZI is composed of two 50:50 beam splitters and two (or one) reconfigurable thermal-optical phase shifters, implementing a unitary operation on two adjacent path modes.
The state preparation module prepares the input herald single photon in an arbitrary two-qubit pure state using three MZIs.  
The measurement module then implements the optimal two-qubit POVMs $\Pi^{(2)}$ using a cascaded MZI network, whose configuration is determined via the algorithm proposed in Ref.~\cite{yan2026single}. The regulation of coefficients $\alpha_{\pm i}$ in $\Pi^{(2)}$ is performed by changing the phases of the on-chip phase shifters. See Supplemental Material for more details.
 
The programmable photonic chip is used to prepare and measure an arbitrary two-qubit pure state. To simulate the measurement results of a mixed two-qubit state, we combine the measurement results from different pure states, each generated with an appropriate probability. 
Specifically, we choose the four pure states as the four orthogonal eigenstates of $\rho_\theta^{\otimes 2}$ with $\theta_x=\theta_y=\theta_z$. All two-copy states under this parameter condition share the same eigenstates, with two of them being degenerate (see Supplemental Material for the explicit forms). 
We spend the same amount of the integral time on each state to simulate the maximally-mixed state (with $\theta_x=\theta_y=\theta_z=0$). 
An estimate of $\theta_{x(y,z)}$ is derived from a dataset of approximately 309 photon detection events. The estimation procedure is repeated 1000 times. Subsequently, the experimental normalized mean squared error for each parameter is calculated by
\begin{equation}
	V_i^\text{exp} = \frac{2N}{1000}\sum_{k=1}^{1000}(\hat\theta_{i,k}-\theta_i)^2, \ i=x,y,z, \label{eq:mse}
\end{equation}
where $\hat\theta_{i,k}$ is the $k$th estimate of $\theta_i$ and $N$ is the average number of photons used for each estimate. The factor of $2$ is included because each photon encodes two qubits. To demonstrate the trade-off relation, we choose 25 weight matrices $W= \operatorname{Diag}(u_x^2,u_y^2,u_z^2)/(u_x^2+u_y^2+u_z^2)$ where each of $u_x$, $u_y$ and $u_z$ takes values traversing the set $\{1,2,3\}$.

Figure~\ref{fig:results_0}(a) demonstrates the three-parameter trade-off relation and the advantage conferred by performing collective two-copy measurements. The experimental mean squared errors (red dots) coincide well with the theoretical predictions (black dots) on the two-copy trade-off surface.   
Figure~\ref{fig:results_0}(b) shows the experimental weighted trace of the normalized mean squared error matrix, that shows the overall error in estimating all three parameters $\theta_{x,y,z}$ simultaneously. The results nearly saturate the two-copy lower bound $\mathcal{C}^{(2)}$ and exceed the single-copy lower bound $\mathcal{C}^{(1)}$ by an average of 16 standard deviations, unequivocally verifying the power of the optimal collective measurement.

\begin{figure*}
	\centering
	\includegraphics[width=\linewidth]{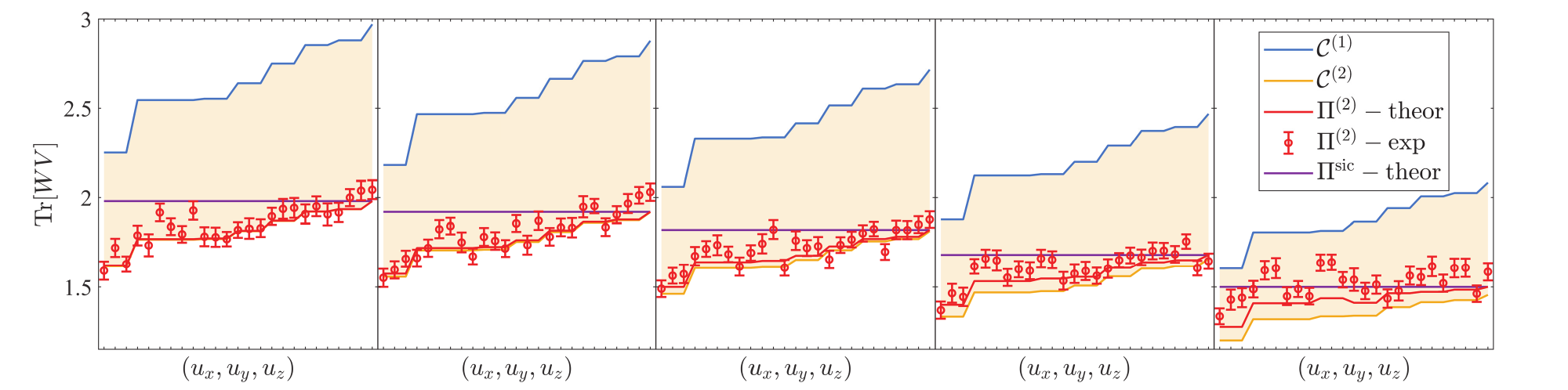}
	\caption{Weighted mean squared error traces for non-optimized parameter values. Left-to-right, the parameters are set to $\theta_x=\theta_y=\theta_z=\theta$ with $\theta=0.1,0.2,0.3,0.4,0.5$, and the average number of photons used for each estimate are $N=309,238,196,156,131$.
    The theoretical performance of the measurement $\Pi^{(2)}$ (red) is close to the two-copy lower bound (yellow). The experimental mean squared errors (red points) are close to their theoretical values, and outperform the two-copy SIC-POVM (purple) over a large range of weights and parameter values. The experiment clearly exceeds the single-copy lower bound (blue). The horizontal axis is the same as in Fig.~\ref{fig:results_0}(b). The error bars denote the standard deviations of 50 numerical simulations from Poisson statistics.}
	\label{fig:weightedMSEother}
\end{figure*}

\textit{Performance for non-maximally-mixed states}---The measurement $\Pi^{(2)}$ is optimal for two-copy tomography around $\theta_0=(0,0,0)$---it is impossible to obtain a lower mean squared error. For practical use in quantum information tasks, its performance for other states is also important. For local estimation at other values of $\theta_0$, $\Pi^{(2)}$ is not optimal, but we nevertheless find that it can still be close to optimal. In Fig.~\ref{fig:weightedMSEother}, we compare it (red) to the NHCRB (yellow) for states with local values $\theta_0=(\theta,\theta,\theta)$ for $\theta >0$. These states are chosen such that experimental data can be obtained from the previous experimental results by post-processing with different proportions of the eigenstates (see Supplemental Material). We see that as the purity of the state increases (i.e., increasing $\theta$), the optimality of $\Pi^{(2)}$ decreases. 

We compare $\Pi^{(2)}$ to the two-copy symmetric informationally complete POVM (SIC-POVM), $\Pi^\text{sic}$, that minimizes the average error given a uniform prior distribution for the parameters~\cite{zhu_universally_2018} and has been demonstrated in photonics~\cite{hou_deterministic_2018} (see End Matter for its definition). This serves as a benchmark to study the improvement that can be gained by optimizing for a particular weight. As shown in Fig.~\ref{fig:results_0}(b) and Fig.~\ref{fig:weightedMSEother}, $\Pi^{(2)}$ can outperform $\Pi^\text{sic}$ when the weights are not equal. This is not surprising given that $\Pi^{(2)}$ is optimized, but the difference highlights that some performance must be sacrificed to perform well across different states and weights.

\textit{Discussion and conclusions}---In summary, we proposed a tight trade-off relation that determines the extractable information for quantum three parameter estimation. We have demonstrated that applying two-parameter trade-off relations, for which there exist general results, to three-parameter estimation problems is not sufficient to obtain a tight trade-off relation. 
However, determining trade-offs from Cram\'er--Rao bounds is a promising alternative and can be applied to any number of parameters. 

We derived three-parameter trade-off relations for qubit tomography based on the NHCRB, finding that the trade-off can be partly mitigated by using collective measurements. 
We showed that these trade-off surfaces are attainable by explicitly determining optimal measurements. These measurements are locally optimal with respect to any measurement on two copies of the state, including those utilizing ancillary resources. Using a programmable photonic chip, we implemented two-copy measurements saturating the trade-off relation. In doing so, we experimentally violate the trade-off relation for separable measurements, explicitly demonstrating the utility of entanglement to mitigate measurement incompatibility. 

Beyond their fundamental interest, trade-off relations are practically important in parameter estimation. Even in full tomography, assigning unequal weights to parameters can be well motivated. For instance, a sensing task may target a primary parameter encoded in one Bloch-sphere projection, with the others carrying only ancillary information. Similarly, in many systems nuisance parameters arise from noise or other processes~\cite{suzuki_nuisance_2020,suzuki_quantum_2020,yung_saturating_2025}. Our results provide useful tools for analyzing such multiparameter scenarios and guiding optimal experimental design.


\textit{Acknowledgements}---This research is supported by the Australian Research Council Centre of Excellence CE170100012 and the National Research Foundation of Korea (RS-2024-00509800). The work at the University of Science and Technology of China is supported by the Quantum Science and Technology-National Science and Technology Major Project (Grant Nos. 2024ZD0300900, 2021ZD0303200, and 2021ZD0301500), the National Natural Science Foundation of China (Grants Nos. 92576203, 92576107, 12534016, T2325022, U23A2074, and 12134014, U25A60082), CAS Project for Young Scientists in Basic Research (YSBR-049 and YSBR-131) and the Anhui Provincial Natural Science Foundation (Grant No.2508085JX001).

\bibliography{bib-1.bib}
\pagebreak
\appendix
\onecolumngrid
\section{\large End Matter}
\subsection*{Optimal two-copy POVM}
Here, we detail the two-copy POVM that is optimal for weighted tomography of the maximally mixed state. Let $\ket{jk_i}:= \ket{j_i}\otimes \ket{k_i}$. Then, define
\begin{align}
	\begin{split}
		\ket{\phi_{+i}} &= \frac{1}{2}\left(\alpha_{+i}\ket{00_i} + \alpha_{-i} \ket{11_i}\right), \\
		\ket{\phi_{-i}} &= \frac{1}{2}\left(\alpha_{-i}\ket{00_i} + \alpha_{+i} \ket{11_i}\right), \\
		\ket{\Psi^-} &= \frac{1}{\sqrt{2}}\left( \ket{01_z}-\ket{10_z} \right)\\
		\alpha_{\pm x} &= \sqrt{\frac{\sqrt{w_x}}{\sqrt{w_x}+\sqrt{w_z}}}\pm\sqrt{\frac{\sqrt{w_x}}{\sqrt{w_x}+\sqrt{w_y}}} \\
		\alpha_{\pm y} &= \sqrt{\frac{\sqrt{w_y}}{\sqrt{w_y}+\sqrt{w_z}}}\pm\sqrt{\frac{\sqrt{w_y}}{\sqrt{w_x}+\sqrt{w_y}}} \\
		\alpha_{\pm z} &= \sqrt{\frac{\sqrt{w_z}}{\sqrt{w_x}+\sqrt{w_z}}}\pm\sqrt{\frac{\sqrt{w_z}}{\sqrt{w_y}+\sqrt{w_z}}} \\
	\end{split} \label{eq:optpovm}
\end{align}
The POVM $\Pi^{(2)} = \{\ketbra{\phi_{+i}},\ketbra{\phi_{-i}},\ketbra{\Psi^-}; \ i=x,y,z\}$ is optimal because it saturates Eq.~\eqref{eq:NH2} and Eq.~\eqref{eq:tradeoff2}. To show this, we consider the outcome probabilities, $p_j = \operatorname{Tr}[\rho^{\otimes 2}\Pi^{(2)}_j]$. To first-order in $\theta_x,\theta_y,\theta_z$ around $(\theta_x,\theta_y,\theta_z)=(0,0,0)$, the outcome probabilties are
\begin{align}
    p_{+x} &= \frac{2w_x+\sqrt{w_xw_y}+\sqrt{w_x w_z}}{8(w_x+\sqrt{w_xw_y}+\sqrt{w_xw_z}+\sqrt{w_yw_z}} + \frac{1}{2}\sqrt{\frac{w_x}{w_x+\sqrt{w_xw_y}+\sqrt{w_xw_z}+\sqrt{w_yw_z}}}\theta_x, \\
    p_{-x} &= \frac{2w_x+\sqrt{w_xw_y}+\sqrt{w_x w_z}}{8(w_x+\sqrt{w_xw_y}+\sqrt{w_xw_z}+\sqrt{w_yw_z}} - \frac{1}{2}\sqrt{\frac{w_x}{w_x+\sqrt{w_xw_y}+\sqrt{w_xw_z}+\sqrt{w_yw_z}}}\theta_x, \\
    p_{+y}  &= \frac{2w_y+\sqrt{w_xw_y}+\sqrt{w_y w_z}}{8(w_y+\sqrt{w_xw_y}+\sqrt{w_xw_z}+\sqrt{w_yw_z}} + \frac{1}{2}\sqrt{\frac{w_y}{w_y+\sqrt{w_xw_y}+\sqrt{w_xw_z}+\sqrt{w_yw_z}}}\theta_y, \\
    p_{-y}  &= \frac{2w_y+\sqrt{w_xw_y}+\sqrt{w_y w_z}}{8(w_y+\sqrt{w_xw_y}+\sqrt{w_xw_z}+\sqrt{w_yw_z}} - \frac{1}{2}\sqrt{\frac{w_y}{w_y+\sqrt{w_xw_y}+\sqrt{w_xw_z}+\sqrt{w_yw_z}}}\theta_y, \\
    p_{+z} &= \frac{2w_z+\sqrt{w_xw_z}+\sqrt{w_y w_z}}{8(w_z+\sqrt{w_xw_y}+\sqrt{w_xw_z}+\sqrt{w_yw_z}} + \frac{1}{2}\sqrt{\frac{w_z}{w_z+\sqrt{w_xw_y}+\sqrt{w_xw_z}+\sqrt{w_yw_z}}}\theta_z,\\
    p_{-z} &= \frac{2w_z+\sqrt{w_xw_z}+\sqrt{w_y w_z}}{8(w_z+\sqrt{w_xw_y}+\sqrt{w_xw_z}+\sqrt{w_yw_z}} - \frac{1}{2}\sqrt{\frac{w_z}{w_z+\sqrt{w_xw_y}+\sqrt{w_xw_z}+\sqrt{w_yw_z}}}\theta_z,\\
    p_\Psi &= \frac{1}{4}.
\end{align}
This leads to the classical Fisher information
\begin{equation}
    F(\Pi^{(2)}) = \operatorname{Diag}\left(\frac{4\sqrt{w_x}}{2\sqrt{w_x}+\sqrt{w_y}+\sqrt{w_z}},\frac{4\sqrt{w_y}}{\sqrt{w_x}+2\sqrt{w_y}+\sqrt{w_z}},\frac{4\sqrt{w_z}}{\sqrt{w_x}+\sqrt{w_y}+2\sqrt{w_z}}\right)
\end{equation}
and mean squared error matrix
\begin{equation}
    V(\Pi^{(2)}) = \operatorname{Diag}\left(\frac{2\sqrt{w_x}+\sqrt{w_y}+\sqrt{w_z}}{4\sqrt{w_x}},\frac{\sqrt{w_x}+2\sqrt{w_y}+\sqrt{w_z}}{4\sqrt{w_y}},\frac{\sqrt{w_x}+\sqrt{w_y}+2\sqrt{w_z}}{4\sqrt{w_z}}\right).
\end{equation}

\subsection*{Two-copy SIC-POVM definition}
The two-copy symmetric informationally complete POVM (SIC-POVM) is a 5-outcome measurement consisting of the elements $
	\Pi^\text{sic}_j = \frac{3}{4}\left(\ketbra{\psi_j}\right)^{\otimes 2}, 
    \Pi^\text{sic}_5 = \ketbra{\Psi^{-}},
$
where $\{\ket{\psi_j}\}_{j=1}^4$ form a single-qubit SIC-POVM (see, e.g., Ref.~\cite{renes_symmetric_2004}). An example is 
\begin{equation}
    \ket{\psi_1} = \ket{0}, \ \ket{\psi_2} = \frac{1}{\sqrt{3}}\left(\ket{0}+\sqrt{2}\ket{1}\right), \ \ket{\psi_3} = \frac{1}{\sqrt{3}}\left(\ket{0}+e^{\frac{2\pi}{3}i}\sqrt{2}\ket{1}\right), \ \ket{\psi_4} = \frac{1}{\sqrt{3}}\left(\ket{0}+e^{-\frac{2\pi}{3}i}\sqrt{2}\ket{1}\right).
\end{equation}
It minimizes the average error given a uniform prior distribution for the quantum state~\cite{zhu_universally_2018}, the opposite extreme to the local assumption. 

\end{document}


\title{Supplementary Material for \\
\textit{Beating three-parameter precision trade-offs with entangling collective measurements}}

\date{\today}

\author{Simon K. Yung}
\thanks{These authors contributed equally to this work}
\affiliation{Department of Quantum Science and Technology, Research School of Physics, The Australian National University, Canberra, ACT 2601, Australia.}

\author{Wen-Zhe Yan}
\thanks{These authors contributed equally to this work}
\affiliation{Laboratory of Quantum Information, University of Science and Technology of China, Hefei, 230026, China}
\affiliation{Anhui Province Key Laboratory of Quantum Network, University of Science and Technology of China, Hefei, 230026, China}

\author{Lan-Tian Feng}
\email{fenglt@ustc.edu.cn}
\affiliation{Laboratory of Quantum Information, University of Science and Technology of China, Hefei, 230026, China}
\affiliation{Anhui Province Key Laboratory of Quantum Network, University of Science and Technology of China, Hefei, 230026, China}
\affiliation{CAS Center For Excellence in Quantum Information and Quantum Physics, University of Science and Technology of China, Hefei, 230026,  China}
\affiliation{Hefei National Laboratory, Hefei, 230088, China}

\author{Aritra Das}
\affiliation{Department of Quantum Science and Technology, Research School of Physics, The Australian National University, Canberra, ACT 2601, Australia.}

\author{Jiayi Qin}
\affiliation{Department of Quantum Science and Technology, Research School of Physics, The Australian National University, Canberra, ACT 2601, Australia.}

\author{Guang-Can Guo}
\affiliation{Laboratory of Quantum Information, University of Science and Technology of China, Hefei, 230026, China}
\affiliation{Anhui Province Key Laboratory of Quantum Network, University of Science and Technology of China, Hefei, 230026, China}
\affiliation{CAS Center For Excellence in Quantum Information and Quantum Physics, University of Science and Technology of China, Hefei, 230026,  China}
\affiliation{Hefei National Laboratory, Hefei, 230088, China}

\author{Ping Koy Lam}
\affiliation{A*STAR Quantum Innovation Centre (Q.INC), Agency for Science, Technology and Research (A*STAR), 2 Fusionopolis Way, Innovis, 138634, Singapore.}
\affiliation{Centre for Quantum Technologies, National University of Singapore, 3 Science Drive 2, Singapore 117543, Singapore.}

\author{Jie Zhao}
\email{jie.zhao@anu.edu.au}
\affiliation{Department of Quantum Science and Technology, Research School of Physics, The Australian National University, Canberra, ACT 2601, Australia.}

\author{Zhibo Hou}
\email{houzhibo@ustc.edu.cn}
\affiliation{Laboratory of Quantum Information, University of Science and Technology of China, Hefei, 230026, China}
\affiliation{Anhui Province Key Laboratory of Quantum Network, University of Science and Technology of China, Hefei, 230026, China}
\affiliation{CAS Center For Excellence in Quantum Information and Quantum Physics, University of Science and Technology of China, Hefei, 230026,  China}
\affiliation{Hefei National Laboratory, Hefei, 230088, China}

\author{Lorc\'an O. Conlon}
\affiliation{A*STAR Quantum Innovation Centre (Q.INC), Agency for Science, Technology and Research (A*STAR), 2 Fusionopolis Way, Innovis, 138634, Singapore.}
\affiliation{Centre for Quantum Technologies, National University of Singapore, 3 Science Drive 2, Singapore 117543, Singapore.}

\author{Syed M. Assad}
\email{cqtsma@gmail.com}
\affiliation{A*STAR Quantum Innovation Centre (Q.INC), Agency for Science, Technology and Research (A*STAR), 2 Fusionopolis Way, Innovis, 138634, Singapore.}

\author{Xi-Feng Ren}
\affiliation{Laboratory of Quantum Information, University of Science and Technology of China, Hefei, 230026, China}
\affiliation{Anhui Province Key Laboratory of Quantum Network, University of Science and Technology of China, Hefei, 230026, China}
\affiliation{CAS Center For Excellence in Quantum Information and Quantum Physics, University of Science and Technology of China, Hefei, 230026,  China}
\affiliation{Hefei National Laboratory, Hefei, 230088, China}

\author{Guo-Yong Xiang}
\affiliation{Laboratory of Quantum Information, University of Science and Technology of China, Hefei, 230026, China}
\affiliation{Anhui Province Key Laboratory of Quantum Network, University of Science and Technology of China, Hefei, 230026, China}
\affiliation{CAS Center For Excellence in Quantum Information and Quantum Physics, University of Science and Technology of China, Hefei, 230026,  China}
\affiliation{Hefei National Laboratory, Hefei, 230088, China}

\maketitle

\section{Preliminaries}
For qubit tomography, we have 
\begin{equation}
	\rho(\btheta) = \frac{1}{2}(\openone_2 + \btheta \cdot \boldsymbol{\sigma}),
\end{equation} 
and 
\begin{equation}
	\partial_i\rho = \frac{1}{2}\sigma_i. 
\end{equation}
The SLD operators are then
\begin{align}
	L_x &= \frac{1}{1-\theta_x^2-\theta_y^2-\theta_z^2}\begin{pmatrix}
		\theta_x(\theta_z-1) & 1-\theta_y^2-\theta_z^2-i\theta_x\theta_y \\
		1-\theta_y^2-\theta_z^2+i\theta_x\theta_y & -\theta_x(\theta_z+1)
	\end{pmatrix} \\
	L_y &= \frac{1}{1-\theta_x^2-\theta_y^2-\theta_z^2}\begin{pmatrix}
		\theta_y(\theta_z-1) & -i(1-\theta_x^2-\theta_z^2+i\theta_x\theta_y) \\
		i(1-\theta_x^2-\theta_z^2-i\theta_x\theta_y) & -\theta_y(\theta_z+1)
	\end{pmatrix} \\
	L_z &= \frac{1}{1-\theta_x^2-\theta_y^2-\theta_z^2}\begin{pmatrix}
		1-\theta_x^2-\theta_y^2-\theta_z & \theta_z(\theta_x-i\theta_y) \\
		\theta_z(\theta_x+i\theta_y) & -1+\theta_x^2+\theta_y^2-\theta_z
	\end{pmatrix}
\end{align}
The quantum Fisher information is
\begin{equation}
	J = \frac{1}{2}\qtr[\rho\{L_i,L_j\}] = \frac{1}{1-\theta_x^2-\theta_y^2-\theta_z^2}\begin{pmatrix}
		1-\theta_y^2-\theta_z^2 & \theta_x\theta_y & \theta_x\theta_z \\
		\theta_x\theta_y & 1-\theta_x^2-\theta_z^2 & \theta_y\theta_z \\
		\theta_x\theta_z & \theta_y\theta_z & 1-\theta_x^2-\theta_y^2 
	\end{pmatrix}.
\end{equation}

\section{Pairwise trade-off relation}

The quantum Fisher information at $\theta_x=\theta_y=\theta_z=0$ is the identity, $J = I_3$. Similarly, fixing one of the parameters (e.g., treating it is a nuisance parameter \cite{suzuki_nuisance_2020,suzuki_quantum_2020}) and considering a two-parameter model, we get $J^{(2)} = I_2$ (because $J$ is diagonal, so there are no correlations between the parameter estimates). Then, the trade-off relation for the mean squared errors of the two-parameter model is \cite{yung_comparison_2024}
\begin{equation}
	(V_i-[J^{-1}]_{ii})(V_j-[J^{-1}]_{jj})\geq 1/\det[J] \Longrightarrow (V_i-1)(V_j-1)\geq 1. 
\end{equation}

\section{Cram\'er--Rao bound calculations}
\subsection{Nagaoka--Hayashi Cram\'er--Rao bound}
The Nagaoka--Hayashi Cram\'er--Rao bound (NHCRB) is \cite{conlon_efficient_2021}
\begin{equation}
	\mathcal{C}_\text{NH} = \min_{\mathbb{L},X}\{\ttr[(W\otimes\rho)\mathbb{L}] \ | \ \mathbb{L}_{jk} = \mathbb{L}_{kj} \text{ Hermitian, } \mathbb{L}\succeq XX^\top \},
\end{equation}
where $X=(X_1,X_2,X_3)^\top$ is a vector of Hermitian operators satisfying the locally unbiased conditions
\begin{equation}
	\qtr[\rho_\btheta X_i]=\theta_i, \quad \partial_j\qtr[\rho_\btheta X_i]=\delta_{ij},
\end{equation}
and we assume $W = \operatorname{Diag}(w_x,w_y,w_z)$. 

In this case, the locally unbiased conditions completely define $X_i = \sigma_i$. Because $\rho = \openone_2/2$ at the true value $\btheta = 0$, only the diagonal elements of $\mathbb{L}$ contribute to the trace and are therefore relevant. We find the following optimal $\mathbb{L}$ based on the numerical solution to the semidefinite program:

\begin{multline}
	\mathbb{L}^\star = \operatorname{Diag}\left(1+\frac{\sqrt{w_y}+\sqrt{w_z}}{\sqrt{w_x}},1+\frac{\sqrt{w_y}+\sqrt{w_z}}{\sqrt{w_x}},1+\frac{\sqrt{w_x}+\sqrt{w_z}}{\sqrt{w_y}},\right.\\ \left. 1+\frac{\sqrt{w_x}+\sqrt{w_z}}{\sqrt{w_y}},1+\frac{\sqrt{w_x}+\sqrt{w_y}}{\sqrt{w_z}},1+\frac{\sqrt{w_x}+\sqrt{w_y}}{\sqrt{w_z}}\right)
\end{multline}
leading to the lower bound 
\begin{equation}
	\mathcal{C}_\text{NH}^{(1)}(W) = (\sqrt{w_x}+\sqrt{w_y}+\sqrt{w_z})^2.  
\end{equation}

Similarly, for the two-copy problem, only the diagonal elements of $\mathbb{L}$ contribute. Furthermore, only the specific sums of the diagonal elements matter. That is,
\begin{equation}
	\ttr[(W\otimes(\openone_4/4)\mathbb{L}] = \frac{1}{4}\left(w_x\underbrace{\sum_{i=1}^4 \mathbb{L}_{ii}}_{\alpha} + w_y\underbrace{\sum_{i=5}^8 \mathbb{L}_{ii}}_{\beta}+ w_z\underbrace{\sum_{i=9}^{12} \mathbb{L}_{ii}}_{\gamma}\right).
\end{equation}
 We find the optimal solutions
\begin{equation}
	\alpha = 2+\frac{\sqrt{w_y}+\sqrt{w_z}}{\sqrt{w_x}}, \quad \beta = 2+\frac{\sqrt{w_x}+\sqrt{w_z}}{\sqrt{w_y}}, \quad \gamma = 2+\frac{\sqrt{w_x}+\sqrt{w_y}}{\sqrt{w_z}},
\end{equation}
which leads to the lower bound
\begin{equation}
	\mathcal{C}_\text{NH}^{(2)} = \frac{1}{2}\left(w_x+w_y+w_z+\sqrt{w_xw_y}+\sqrt{w_xw_z}+\sqrt{w_yw_z}\right).
\end{equation}

\subsection{Holevo Cram\'er--Rao bound}
The asymptotic limit that applies to measurements performed jointly on infinitely many copies of the state is given by the Holevo Cram\'er--Rao bound~\cite{nagaoka_new_2005}:

\begin{equation}
    \mathcal{C}_\text{H} = \min_X\left\{ \Tr[W\re  Z_\btheta[X]]+\operatorname{TrAbs}[W \operatorname{Im} Z_\btheta[X]] \right\},
\end{equation}
where $Z_\btheta[X]$ is the matrix with elements $Z_\btheta[X]_{ij}=\Tr[\rho_\btheta X_iX_j]$, and $X_i$ obey the locally unbiased conditions. As before, $X_i$ are constrained to $X_i = \sigma_i$, and therefore $Z_\btheta[X]=\openone_3$. Hence, $\mathcal{C}_H = \Tr[W]$, which is equal to the quantum Cram\'er--Rao bound. That is, in the asymptotic limit, the trade-off can be entirely eliminated. 

\section{Trade-off relation calculations}\label{sec:tradeoff}
To determine the trade-off region for single-copy measurements, we consider the planes defined by the weighted NHCRB:
\begin{equation}
	w_x V_x + w_y V_y + w_z V_z \geq (\sqrt{w_x}+\sqrt{w_y}+\sqrt{w_z})^2,
\end{equation}
for different values of $w_x,w_y,w_z$. The combination of all such planes determines the overall accessible region. To determine the boundary of this region, we consider parameterizing the weights by
\begin{equation}
	w_x = s^2,\quad w_y = t^2, \quad w_z = (1-s-t)^2. 
\end{equation}
These are chosen based on the form of $\mathcal{C}_\text{NH}^{(1)}$ (Note that the weights need not have a constant sum---it is their ratio that controls the halfspace). The boundary is determined by eliminating $s$ and $t$ from the following simultaneous equations:
\begin{align}
	s^2 V_x + t^2 V_y + (1-s-t)^2 V_z &= \mathcal{C}_\text{NH}^{(1)}(s^2,t^2,(1-s-t)^2), \\
	\frac{\partial}{\partial s}\left(s^2 V_x + t^2 V_y + (1-s-t)^2 V_z\right) & = \frac{\partial}{\partial t}\mathcal{C}_\text{NH}^{(1)}(s^2,t^2,(1-s-t)^2), \\
	\frac{\partial}{\partial t}\left(s^2 V_x + t^2 V_y + (1-s-t)^2 V_z\right) & = \frac{\partial}{\partial t}\mathcal{C}_\text{NH}^{(1)}(s^2,t^2,(1-s-t)^2).
\end{align}
Substituting $\mathcal{C}_\text{NH}^{(1)}$, we have
\begin{align}
    s^2 V_x + t^2 V_y + (1-s-t)^2 V_z &= (s+t+(1-s-t))^2 = 1, \label{eq:1}\\
    2s V_x -2(1-s-t)V_z &= 0 \Longrightarrow s V_x =(1-s-t)V_z, \label{eq:2}\\
    2t V_y -2(1-s-t)V_z &= 0\Longrightarrow t V_y = (1-s-t)V_z. \label{eq:3}\\
    &
\end{align}
Equations \eqref{eq:2} and \eqref{eq:3} can be rearranged to $sV_x=tV_y$ and $(1-s-t)=sV_x/V_z$. Then, we can solve for $s$ and $t$:
\begin{align}
    \left(1-s-\left(s V_x/V_y\right)\right) = sV_x/V_z \Longrightarrow s &= \frac{V_yV_z}{V_xV_y+V_xV_z+V_yV_z}, \\
    \Longrightarrow t &= \frac{V_xV_z}{V_xV_y+V_xV_z+V_yV_z}.
\end{align}
Then 
\begin{equation}
    1-s-t = 1-\frac{V_yV_z}{V_xV_y+V_xV_z+V_yV_z}-\frac{V_xV_z}{V_xV_y+V_xV_z+V_yV_z} = \frac{V_xV_y}{V_xV_y+V_xV_z+V_yV_z}.
\end{equation}
Substituting into Eq.~\eqref{eq:1}, we get
\begin{equation}
    \frac{(V_yV_z)^2V_x + (V_xV_z)^2V_y+(V_xV_y)^2V_z}{(V_xV_y+V_xV_z+V_yV_z)^2} = 1\Longrightarrow V_xV_yV_z \frac{V_yV_z+V_xV_z +V_xV_y}{(V_xV_y+V_xV_z+V_yV_z)^2}=1,
\end{equation}
which gives 
\begin{equation}
	V_xV_yV_z -V_xV_y-V_xV_z-V_yV_z=0. 
\end{equation}
The conditions $V_i\geq 1$ are also necessary to satisfy the quantum Cram\'er--Rao bound for each parameter. 

The trade-off region for two-copy measurements can be calculated using the same approach, replacing $\mathcal{C}_\text{NH}^{(1)}$ with $\mathcal{C}_\text{NH}^{(2)}$.

\section{Single-Copy Optimal POVMs}

Here, we present an optimal single-copy POVM for weighted qubit tomography of the maximally mixed state. Let $\ket{j_i}$, $i\in \{x,y,z\}$, $j\in \{0,1\}$ be the normalized eigenvectors of $\sigma_i$. Then, define
\begin{align}\label{eq:optpovm}
	\begin{split}
	\ket{\psi_1} &= a_x \ket{0_x}, \ \ket{\psi_2} = a_x \ket{1_x},
	\ket{\psi_3} = a_y \ket{0_y}, \ \ket{\psi_4} = a_y \ket{1_y},
	\ket{\psi_5} = a_z \ket{0_z}, \ \ket{\psi_6} = a_z \ket{1_z}, \\
	a_i &= \sqrt{\frac{\sqrt{w_i}}{\sqrt{w_x}+\sqrt{w_y}+\sqrt{w_z}}}.
	\end{split}
\end{align}
The POVM $\Pi^{(1)} = \{\ketbra{\psi_i},\ 1\leq i\leq 6\}$ is optimal. The form of this measurement is intuitive: each pair of elements ($\{1,2\}$, $\{3,4\}$, $\{5,6\}$) provides information about the projection of the Bloch vector along the respective axis (i.e., one of the parameters), but no information about the others. The factors $a_i$ apportion the detection probabilities based on the relative weights. 

The outcome probabilities are
\begin{align}
	p_i &= \qtr[\rho\ketbra{\psi_i}] = \qtr\left[\frac{1}{2}(\openone_2+\btheta\cdot \boldsymbol{\sigma})\ketbra{\psi_i}\right] = \frac{1}{2}\bra{\psi_i}(\openone + \btheta \cdot\boldsymbol{\sigma})\ket{\psi_i}	\\
	p_1 &= \frac{1+\theta_x}{2}a_x^2, \quad p_2 = \frac{1-\theta_x}{2}a_x^2 \\
	p_3 &= \frac{1+\theta_y}{2}a_y^2, \quad p_4 = \frac{1-\theta_y}{2}a_y^2 \\
	p_5 &= \frac{1+\theta_z}{2}a_z^2, \quad p_6 = \frac{1-\theta_z}{2}a_z^2 \\
\end{align}
From these probabilities, we calculate the classical Fisher information (about $\theta_x=\theta_y=\theta_z=0$) as
\begin{equation}
	F(\Pi^{(1)})= \begin{pmatrix}
		a_x^2 & 0 & 0 \\
		0 & a_y^2 & 0 \\
		0 & 0 & a_z^2
	\end{pmatrix} = \begin{pmatrix}
 	\frac{\sqrt{w_x}}{\sqrt{w_x}+\sqrt{w_y}+\sqrt{w_z}} & 0 & 0 \\
		0 & \frac{\sqrt{w_y}}{\sqrt{w_x}+\sqrt{w_y}+\sqrt{w_z}} & 0 \\
		0 & 0 & \frac{\sqrt{w_z}}{\sqrt{w_x}+\sqrt{w_y}+\sqrt{w_z}}
 \end{pmatrix}
. 
\end{equation}
The mean squared error matrix is thus
\begin{equation}
	V(\Pi^{(1)}) = \begin{pmatrix}
		 \frac{\sqrt{w_x}+\sqrt{w_y}+\sqrt{w_z}}{\sqrt{w_x}} & 0 & 0 \\
		 0 & \frac{\sqrt{w_x}+\sqrt{w_y}+\sqrt{w_z}}{\sqrt{w_y}} & 0 \\
		 0 & 0 & \frac{\sqrt{w_x}+\sqrt{w_y}+\sqrt{w_z}}{\sqrt{w_z}},
	\end{pmatrix}
\end{equation}
which reflects that as a weight is increased, the corresponding mean squared error is reduced at the cost of increasing the mean squared errors for the other parameters.

\section{Equivalent results for non-maximally-mixed probe}
In the main text, we present results for a probe state with $\theta_x=\theta_y=\theta_z=0$. Here, we provide an example of results for a different probe state, demonstrating that equivalent results can be obtained for any local parameter values. 

\begin{figure}
    \centering
    \includegraphics[width=0.5\linewidth]{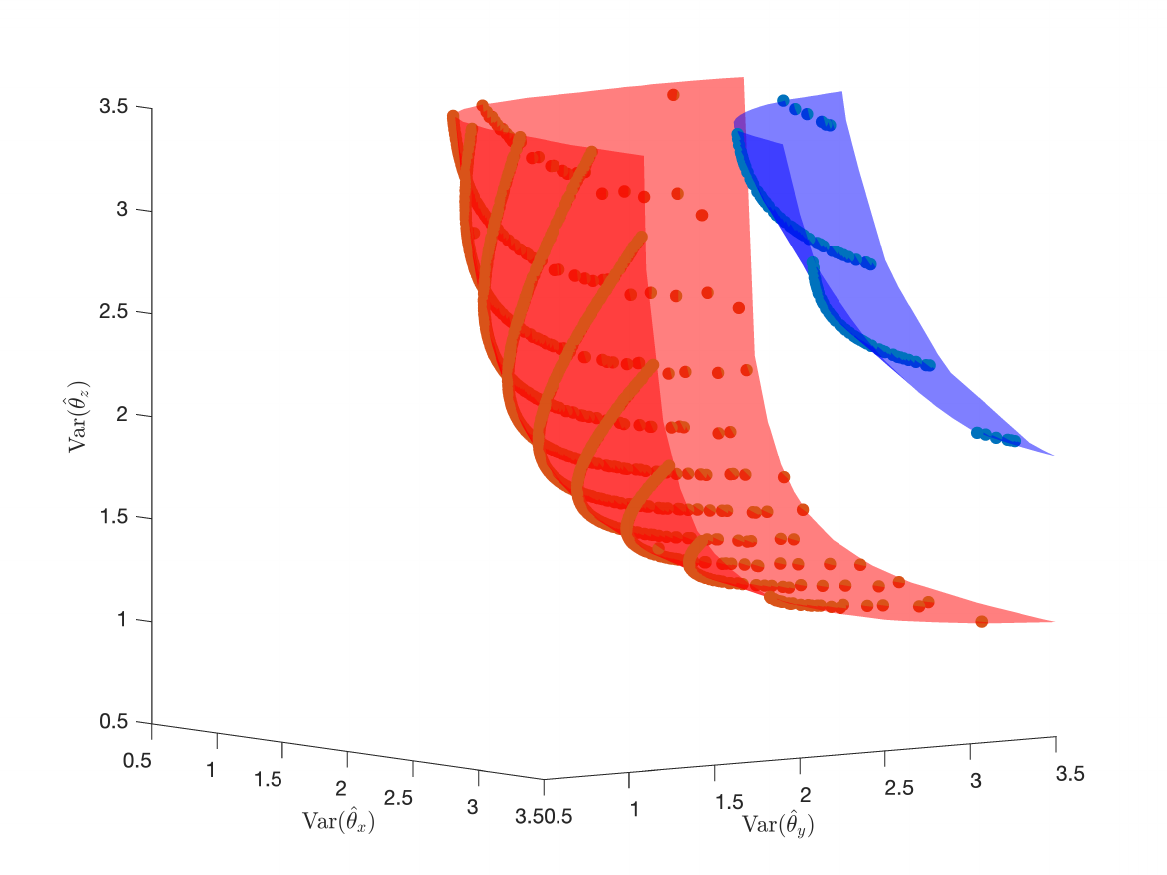}
    \caption{Trade-off surfaces for $(\theta_x,\theta_y,\theta_z)=(0.3,0.3,0.3)$ with one (blue) and two (red) copies. The points on the surfaces are the points of intersection of the planes for different weighted NHCRBs.}
    \label{fig:surfex1}
\end{figure}

Arbitrarily, we choose the probe state with $(\theta_x,\theta_y,\theta_z)=(0.3,0.3,0.3)$. We calculate the weighted NHCRBs numerically using a semidefite program \cite{conlon_efficient_2021} for a range of different weight matrices. We then use these lower bounds to construct an approximate trade-off surface, combining the halfspaces defined by each weighted bound. This is a discretised version of the calculation in Sec.~\ref{sec:tradeoff}, and we use the Quickhull algorithm \cite{barber_quickhull_1996} implemented in the \texttt{qhalf} routine of the \texttt{Qhull} program \footnote{{https://www.qhull.org/html/qhalf.htm}}. This numerical method was previously used in Ref.~\cite{yung_saturating_2025} to determine two-parameter trade-off relations. The resultant trade-off surfaces are presented in Fig.~\ref{fig:surfex1}, including the intersection points of the planes that make up the surface. 

For this probe state and the weight matrix $W = \operatorname{diag}(1^2,2^2,3^2)/(1^2+2^2+3^2)$, we find (non-unique) POVM via numerical optimisation, with optimality confirmed by comparing the classical Fisher information to the NHCRB. For a single copy, the POVM is defined by the projectors:
\begin{align}
   \ket{\pi^{(1)}_1} = \begin{pmatrix}
       0.1743  \\
   0.007- 0.7019i 
    \end{pmatrix}, \ket{\pi^{(1)}_2} = \begin{pmatrix}
        0.4374 \\
        -0.1208+0.5850i 
    \end{pmatrix}, \\ \ket{\pi^{(1)}_3} = \begin{pmatrix}
        0.6709\\
        0.2631-0.0472i
    \end{pmatrix}, \ket{\pi^{(1)}_4} = \begin{pmatrix}
        0.5729\\
        -0.2179-0.1778
    \end{pmatrix}.
\end{align}

Similarly, for two copies the POVM defined by the following projectors is optimal
\begin{align}
        \ket{\pi^{(2)}_1} = \begin{pmatrix}
        0.2806\\
        0.3724-0.0137i \\
        0.3724-0.0137i\\
        0.3168-0.0288i
    \end{pmatrix}, \ket{\pi^{(2)}_2} = \begin{pmatrix}
        0.3097 \\
        -0.0874+0.4903i \\
        -0.0874+0.4903i \\
        -0.3358-0.1722i
    \end{pmatrix}, \ket{\pi^{(2)}_3} = \begin{pmatrix}
        0.1915\\
        0.0467-0.2655i \\
        0.0467-0.2655i \\
        -0.2746-0.0893i
    \end{pmatrix}, \\\ket{\pi^{(2)}_4} = \begin{pmatrix}
        0.8875\\
        -0.0923-0.1078i \\
        -0.0923-0.1078i \\
        -0.2746-0.0893i
    \end{pmatrix}, \ket{\pi^{(2)}_5} = \begin{pmatrix}
        0.0330 \\
        -0.1351-0.0438i \\
        -0.1351-0.0438i \\
        0.1982+0.7909i
    \end{pmatrix},\ket{\pi^{(2)}_6} = \begin{pmatrix}
        0 \\
        0.7071  \\
        -0.7071 \\
        0
    \end{pmatrix}.
\end{align}

\section{Experimental details}
\subsection{Experimental setup}
The integrated herald single-photon source is based on a 1.2 cm long spiral silicon waveguide. We use a femtosecond erbium laser with a wavelength centered at 1550.12 nm and a repetition rate of 100 MHz to pump the waveguide. In the waveguide, the spontaneous four-wave mixing process occurs and correlated photon pairs are generated. At the output of the waveguide, the remaining pump photons are removed by two cascaded post-filters. The signal and idler
photons, which have respective central wavelengths of 1561.42 nm and 1538.98 nm, are selected by two wavelength division multiplexing filters. The idler photons are directly detected by a superconducting nanowire single-photon detector as triggers, whereas the signal photons are input into the subsequent programmable chip to undergo the state preparation and measurement process. 

On the programmable chip shown in Fig.2 of the main text, the quantum states are encoded in the path degree of freedom of a single photon and manipulated by combinations of Mach-Zehnder interferometers (MZIs). 
In the state preparation module, an arbitrary four-dimensional pure state can be prepared using three MZIs. 
This is achieved with three phase shifters (placed between the two beam splitters in each MZI) collectively adjusting the amplitudes across the four path modes, and another three phase shifters (placed after the second beam splitter in each MZI) adjusting the relative phases. 
The measurement module, built from a cascaded MZI network, can implement an arbitrary four-dimensional measurement composed of seven rank-1 operators. 
The on-chip phases are determined via the algorithm proposed in Ref.~\cite{yan2026single}. Details of the chip fabrication and calibration can also be found in Ref~.\cite{yan2026single}.

\subsection{Demonstration of the two-copy state tomography}

\begin{figure*}
	\centering
	\includegraphics[width=0.7\linewidth]{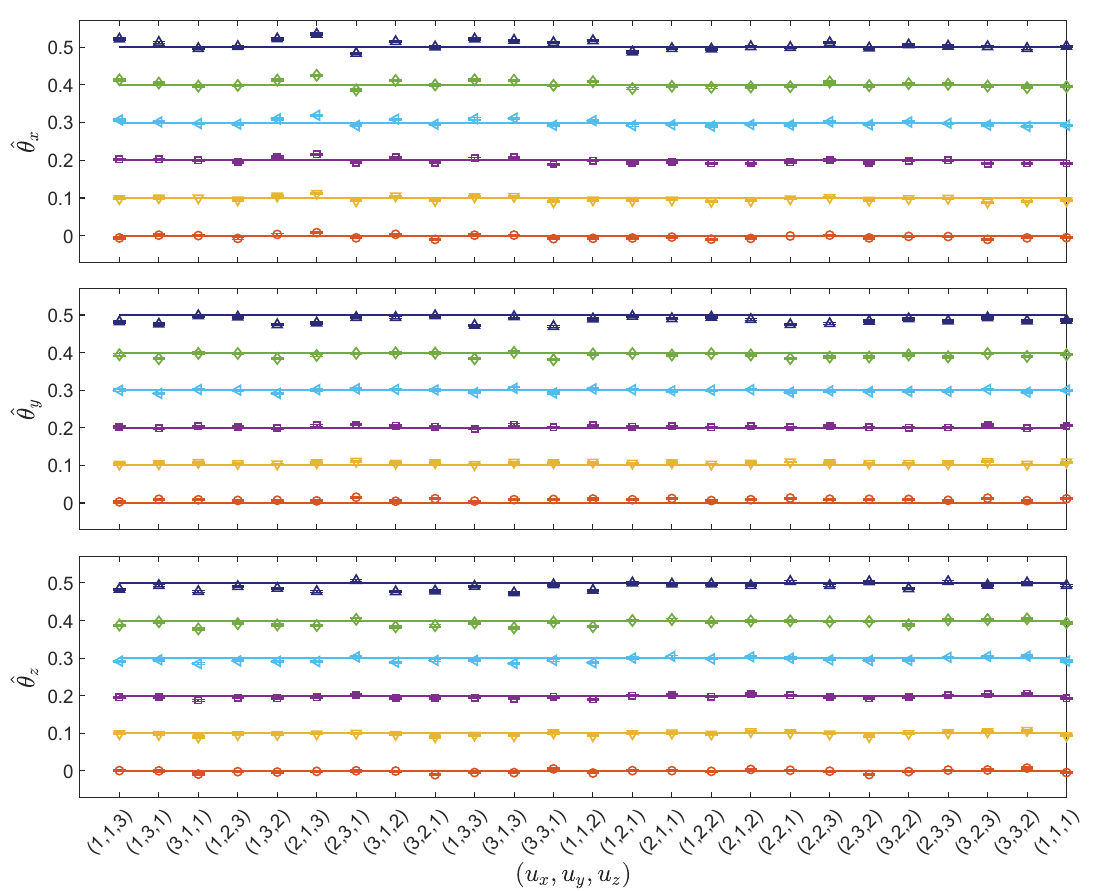}
	\caption{Estimated values of the three parameters with their true values set as
    $\theta_x=\theta_y=\theta_z=0,\ 0.1,\ 0.2,\ 0.3,\ 0.4,\ 0.5$. 
    The horizontal axis corresponds to the weight matrix via $W= \operatorname{Diag}(u_x^2,u_y^2,u_z^2)/(u_x^2+u_y^2+u_z^2)$. Each data point is the average over 1000 repeated experiments, and the error bar indicates the standard deviation. The solid lines represent the theoretical predictions.
    }
	\label{fig:estimate}
\end{figure*}

We experimentally demonstrated quantum state tomography for a class of two-copy states $\rho_\theta^{\otimes 2}$ whose parameters satisfy $\theta_x=\theta_y=\theta_z$.
The tomography is based on the optimal two-copy collective measurement designed for the maximally mixed state [Eq. (6) of the maintext].
Because our chip can only prepare pure four-dimensional states, we performed the measurement on each of the following four states
\begin{align}
	\begin{split}
		\ket{\varphi_1} &= \begin{pmatrix}
			-0.2113i \\
			0.2887(i-1) \\
			0.2887(i-1) \\
			0.7887
		\end{pmatrix},\ \ket{\varphi_2} = \begin{pmatrix}
			-0.4757i\\
			0.1617+0.2093i\\
			-0.6375+0.2665i \\
			-0.4757
		\end{pmatrix}, \\
		\ket{\varphi_3} &= \begin{pmatrix}
			-0.3271i \\
			-0.7447+0.2051i \\
			0.4176+0.1220i \\
			-0.3271
		\end{pmatrix},\ \ket{\varphi_4} = \begin{pmatrix}
			0.7887i \\
			0.2887(i-1) \\
			0.2887(i-1) \\
			-0.2113
		\end{pmatrix},
	\end{split} \label{eq:states}
\end{align}
which are the common eigenstates of $\rho_\theta^{\otimes 2}$ with $\theta_x=\theta_y=\theta_z$. 
The measurement results for this class of two-copy states 
were obtained by summing the collected photon counts for these four eigenstates, whose integral times were set proportional to their corresponding eigenvalues in the target state. 
The adjustment of the integral time for each eigenstate were realized by truncating its raw experimental data (the photon time tags recorded over a sufficiently long acquisition period) to an appropriate length.
In the experiment, we set $\theta_x=\theta_y=\theta_z=0,\ 0.1,\ 0.2,\ 0.3,\ 0.4,\ 0.5$. The integration times for the four eigenstates were set proportional to (0.25,0.25,0.25,0.25), (0.1709,0.2425,0.2425,0.3411), (0.1068, 0.2200, 0.2200, 0.4532), (0.0577, 0.1825, 0.1825, 0.5773), (0.0236, 0.1300, 0.1300, 0.7164), (0.0045, 0.0625, 0.0625, 0.8705), respectively.

According to the experimental measurement results, denoted as $\{N_i\}_{i=1}^7$ where $N_i$ is the photon counts associated with the $i$th POVM element, we reconstructed the parameters of the target two-copy state through an unbiased estimator. specifically, for the maximally mix state with $\theta_x=\theta_y=\theta_z=0$, we reconstructed it through a linear estimator derived from Eqs.(7)-(12) of the main text.
For the other states with $\theta_x=\theta_y=\theta_z>0$, we reconstructed them through the maximum likelihood estimator using the accelerated projected-gradient algorithm as in Ref.~\cite{hou_deterministic_2018}. The experiment
was repeated 1000 times to get the distribution of the reconstructed parameters.
The mean values of the reconstructed parameters $(\hat{\theta}_x, \hat{\theta}_y, \hat{\theta}_z)$ are shown in Fig.~\ref{fig:estimate} for all configurations of true parameters and weight matrices, while the corresponding weighted mean squared errors are presented in Fig.3 and Fig.4 of the main text, respectively.

\bibliography{bib-1.bib}